\documentstyle[prl,aps,epsf,twocolumn]{revtex}
\newcommand{\isum}%
{\mathop{\hbox{$\displaystyle\sum\kern-13.2pt\int\kern1.5pt$}}}

\begin{document}
\preprint{submitted to Physical Review Letters; LB6633}
\draft
\title{Convergence of two-center expansions in positron-hydrogen collisions}
\author{Alisher S. Kadyrov\thanks{electronic address:
Alisher.Kadyrov@flinders.edu.au} and Igor Bray}
\address{
Electronic Structure of Materials Centre,
The Flinders University of South Australia,
G.P.O. Box 2100, Adelaide 5001, Australia}
\date{\today}
\maketitle
\begin{abstract}
The positron-hydrogen atom scattering system is considered within the
S-wave model. Convergence in the elastic scattering, excitation,
ionization, and positronium formation channels is studied as a
function of the number and type of states used to expand the total
wave function. It is found that all unphysical
resonances disappear only if near-complete
pseudostate expansions are applied to both the atomic and positronium
centers.
\end{abstract}
\pacs{34.85.+x}

In recent years there has been substantial progress in the field of
electron-atom collisions. Close-coupling based methods, utilizing
near-complete pseudostate expansions about the atomic center, have
been shown to yield very accurate results for discrete excitation and
ionization channels, see for example
\cite{BS92,KW95,BBBS96,Betal96}. This progress is primarily based on
the study of the
electron-hydrogen atom scattering S-wave model (only states with zero
orbital angular momentum are retained) performed by Bray and
Stelbovics~\cite{BS92l}. They showed, by simply taking a pseudostate
expansion whose completeness improves with increasing number of states
$N$, that cross sections for discrete and ionization channels converged at all
energies. Pseudoresonances, typically
associated with small $N$ calculations, disappeared for sufficiently
large $N$, and convergence
was to the correct independently evaluated results. This was an
extraordinarily powerful result that is at the base of the substantial
success recently enjoyed by the various implementations of the
close-coupling theories.

The situation for positron-atom scattering is somewhat more
complicated. In addition to the atomically centered states the
positronium formation channel needs to be included in the calculations
in order to allow for all possible scattering channels. In other
words, one needs a combined basis consisting of two independent basis
sets.  However, in this case if each of the components of the basis is
large enough, one may expect instabilities in the calculations. The
reason for this is that at small distances between colliding fragments
functions of different basis components may essentially repeat each
other due to their nonorthogonality. Thus, the close--coupling problem
with the combined--basis--expansions is ill--conditioned. Is this an
insurmountable obstacle? This question has remained unanswered for a
long time. Mitroy, Berge, and Stelbovics~\cite{MBS94} and Mitroy and
Ratnavelu~\cite{MR95} have performed
convergence studies for the full positron-hydrogen problem at low
energies. Below the first hydrogen excitation
threshold they showed good agreement between large pseudostate
close coupling calculations and the highly accurate variational
calculations of Humberston~\cite{H84}.  
However, at higher energies, particularly
above the ionization threshold the situation is less clear.

We adopt the often-used CC($N,N'$) notation for close-coupling
calculations that utilise $N$ atomic eigenstates and $N'$ positronium
eigenstates to expand the total scattering wave function. In
addition, a bar, when applied to $N$ or $N'$ indicates that
pseudostates rather than eigenstates are used.

In 1991 Higgins and Burke~\cite{HB91} showed that in the
close-coupling calculation CC(1,1) a giant resonance appeared around
40~eV incident energy. Since that time a huge body of literature has
been devoted to the study of this and other resonances above the
ionization threshold, see the excellent review of Walters {\em et
al.}~\cite{Walters97}.  The positions and widths of the resonances
have been studied extensively
\cite{HNB91,SMBG93,HB93,Mitroy93b,MS94L,MS94,KMW94a}, even though, as 
noted by Walters~{\em et al.}~\cite{Walters97}, the mid-seventies work
of Simon~\cite{S74,S78} shows that there may not be any
positron-hydrogen scattering resonances above the ionization
threshold.

As far as we are aware the first convincing numerical evidence that
shows disappearance of the above-threshold Higgins-Burke type
resonances was given by Kernoghan, McAlinden, and
Walters~\cite{KMW94,KMW95}. They considered the full positron-hydrogen
scattering problem using pseudostate expansions on both centers. Their
18-state  CC($\overline{9},\overline{9}$) calculation
included s, p, and d states for both centers. Thus it was not
clear what was the primary reason, if any, for the disappearance of
the Higgins-Burke type resonances.  Moreover, at energies above the
ionization threshold the authors encountered a new false
pseudoresonance structure associated with the unphysical pseudostates.
Smooth results for the total and the
dominant partial $1s$ cross sections were obtained after applying an energy
averaging procedure in order to remove the pseudoresonances. The
unsmoothed partial $2s$ and $2p$ cross sections both for atom
excitation and positronium formation contained significantly more
pseudostructure \cite{KRMW96}. 

At the same time the convergent
close-coupling theory (CCC) \cite{BS94}, which made use of
CC($\overline{N},0$) calculations, i.e.
neglected positronium formation, gave very good results for the total,
elastic, excitation and ionization cross sections in energy regimes
where positronium formation cross section is either zero or small. The
CCC calculations 
showed no pseudoresonances.  Based on these results Kernoghan {\em et
al} \cite{KRMW96} (see also \cite{Mitroy96}) suggested using 
pseudostates only for hydrogen with a few eigenstates of
positronium i.e., CC($\overline{N},N'$) type calculations.  Results of
a CC($\overline{30},3$) calculation with a 
30--state hydrogenic pseudobasis of Bray and Stelbovics~\cite{BS94},
supplemented by the three lowest positronium 
eigenstates showed a considerable improvement over the
CC($\overline{9},\overline{9}$) results~\cite{KMW95}.  Though the new
basis did not completely remove false 
pseudostructure from the $2s$ and $2p$ positronium formation cross
sections in the upper neighbourhood of the ionization threshold, the
conclusion was that in two--centre scattering problems 
CC($\overline{N},N'$) type calculations are adequate.  Since the works
of Kernoghan {\em et
al} \cite{KRMW96} and Mitroy~\cite{Mitroy96}  pseudostate calculations of
positron scattering off atoms have been performed mostly using bases
of this type.  The same conclusion became dominant in ion-atom collisions as
well mainly due to the extensive investigations by Kuang and Lin
\cite{KL96,KL96b,KL97}.

The purpose of the present Letter is to demonstrate a case where stable and
convergent two--centre pseudosate calculations free of any
pseudoresonances could be achieved
only if near-complete pseudostate expansions are applied to both the
atomic and positronium centers, i.e. where calculations of type
CC($\overline{N},\overline{N'}$) are
used. This is done for the simple S-wave model that retains only
states of zero orbital angular momentum.  The importance of the
electron-atom scattering 
S-wave model, which continues to attract
considerable attention~\cite{MKW99,BRIM99,S99l}, suggests that this model
might also be useful in the positron-hydrogen case. However, for this
case it has attracted little attention, with the CC(2,2) calculation
by Mitroy~\cite{Mitroy93b} being the biggest. Yet, this problem
captures most of the difficulties associated with the full
positron-atom problem just as the electron-atom S-wave model contains
most of the difficulties associated with the full electron-atom
scattering problems. At the same time it does not include the unnecessary,
in the present context, generalities. Particularly, there is no {\em a
priori} mechanism for any resonances including the Feshbach ones
below the ionization threshold, due to the absence of states with
non-zero angular momentum. This is why the model is ideal for
convergence studies, as only smooth cross sections are expected. 

The convergent close-coupling method~\cite{BS92} is extended by
including the positronium formation channels. This extension is mainly
based on the work of Mitroy~\cite{Mitroy93b,M93}. However, positronium
formation matrix elements have been written as a coupling of 12
$j$-symbols resulting in only two--dimensional integrals and finite
angular momentum sums. In addition, momentum--space pseudostates and
corresponding formfactors are used in a compact analytical form. To
evaluate the integral over the momentum of the virtual electron
involving the logarithmic singularity special--purpose orthonormal
polynomials have been calculated which yield an optimal Gaussian
quadrature. Together, these features allow big pseudostate
calculations to be performed efficiently. The generalized CCC computer
code has been tested against the CC(3,3) calculations of Mitroy and
Stelbovics~\cite{MS94} and other momentum space close--coupling
calculations.

To evaluate the positron-hydrogen S-wave model
the total scattering wave function is expanded in terms of bases
consisting of two independent truncated Laguerre bases with corresponding
exponential fall-off factors $\lambda$ and $\lambda'$,
leading to 
close-coupling calculations denoted by CC($\overline{N},\overline{N'}$).
The $\overline{N}$ hydrogen (H) states were obtained by diagonalising the
Hamiltonian with $\lambda\approx 1.0$. The minor variation in
$\lambda$ was made to ensure that the total energy $E$ was exactly
half-way between two adjacent pseudothresholds, as the underlying integration 
rule requires~\cite{BS95,BC97}. The positronium (Ps) states were obtained
by diagonalising the corresponding
Hamiltonian with $\lambda'=0.5$. This choice for $\lambda$ and $\lambda'$
results in approximately equal number of negative- and positive-energy states.
No variation of $\lambda'$ was performed due to the fact that at the
larger energies (above 40~eV) 
considered here too large a variation would be necessary to ensure that
$E$ is half-way between two adjacent Ps pseudothresholds. Instead, 
the CC($\overline{N},\overline{N'}$) calculations are performed for
all possible  $E$ that are half-way between two of the Ps
pseudothresholds. Though in this case observables 
are calculated at predefined incident energies, to show
convergence in the expected to be smooth cross sections, we combine the
results for varying $\overline{N}$ and $\overline{N'}$. If convergence 
is obtained at the calculated energies such combined results should form
smooth curves.

To assure the convergence on both basis
components we took $\overline{N}=\overline{N'}$. In general, the
combined basis does not need to be 
symmetric in the number of the hydrogen and positronium pseudostates.  The
basic momentum space integral equations for the transition matrix
elements have been solved using a 96-point Gauss quadrature at each
incident energy.  The accuracy of the solution of the integral
equations has been carefully checked for the case of the largest
basis. Any further increase in the number of quadrature points did not
significantly change the results.
The number of states $\overline{N}=\overline{N'}$  has been systematically
increased from 1 to 17.

\begin{figure}
\hspace{-1.5truecm}\epsfxsize=9.5cm\epsffile{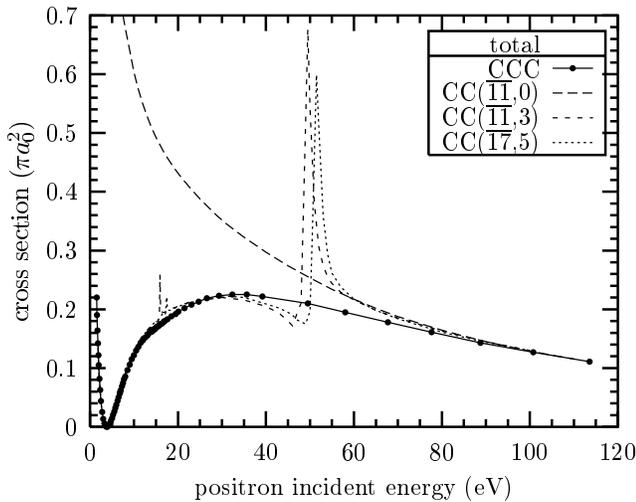}
\caption{Total cross section for the positron--hydrogen scattering
S--wave model. The CCC-labeled curve is obtained from
CC($\overline{N},\overline{N'}$) calculations with
$\overline{N}=\overline{N'}=11,\dots,17$, see text. The
CC($\overline{N},N'$) calculations utilise $\overline{N}$ atomic
pseudostates and $N'$ positronium eigenstates.}
\label{Fig1}
\end{figure}

In Fig.~\ref{Fig1}
 the total cross section for the model is presented evaluated 
using a number of different bases. The CC($\overline{N},N'$)
calculations were performed on a fine energy mesh
allowing for representation as a continuous curve. The dotted curve
labeled CCC is obtained from
all of the CC($\overline{N},\overline{N'}$) calculations (one
for each dot) with $\overline{N}=\overline{N'}=11,\dots,17$, see
above. For example, the results at around 114 and 39~eV are from the
CC($\overline{17},\overline{17}$) calculation and the ones at around
101 and 36~eV are from the CC($\overline{16},\overline{16}$). 

Starting with the
CC($\overline{11},0$) calculation we see a smooth cross section that
is very large at small energies. Adding three Ps eigenstates ($N'=3$) results
in a massive drop in the cross section at low energies (in fact $N'=1$ is 
sufficient for this drop), but leads to Higgins-Burke type resonances
at around 
15 and 50~eV. Adding a further two Ps eigenstates ($N'=5$) and
increasing the atomic 
pseudostate expansion to $\overline{N}=17$ results in only a slight shift
of the resonance to higher energies. The shift is due to the
increasing number of Ps eigenstates as we found invariance in the
CC($\overline{N},3$) cross sections for 
$\overline{N}=11,\dots,17$. However, the CCC curve is smooth devoid of any
resonance structure. Thus, within this model the only practical way to 
yield pseudoresonance-free cross sections is to utilize
near-complete pseudostate expansions for both the atomic and Ps
centers. We note that the elimination of the 
Higgins--Burke type resonances is achieved for relatively small
 $\overline{N}=\overline{N'}=5$, whereas pseudoresonance
structure disappears from $\overline{N}=\overline{N'}=11$.

\begin{figure}
\hspace{-1.5truecm}\epsfxsize=9.5cm\epsffile{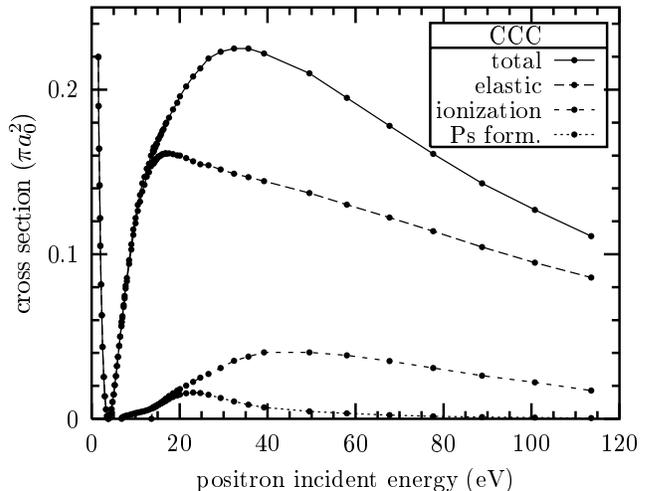}
\caption{Total cross section and its components for the S-wave
model. See text for the description of the CCC calculations.}
\label{Fig2}
\end{figure}

Fig.~\ref{Fig2} shows some of the individual components of the CCC
total cross section given in Fig.~1. The most dominant is the elastic
scattering cross section followed by the ionization and Ps
formation cross sections. The Ps formation cross section is evaluated
by summing the cross sections for the negative-energy Ps states
projected onto the Ps true discrete spectrum. 
The ionization (breakup) cross section is
evaluated by subtracting from the total cross section the Ps formation 
cross section and the sum of the atomic negative-energy state cross sections 
projected onto the true discrete atomic spectrum. Though the Ps
formation cross section is generally very small the 
inclusion of the Ps channels considerably reduces the total cross section
at energies below 70~eV (see Fig.~\ref{Fig1}).

\begin{figure}
\hspace{-1.5truecm}\epsfxsize=9.5cm\epsffile{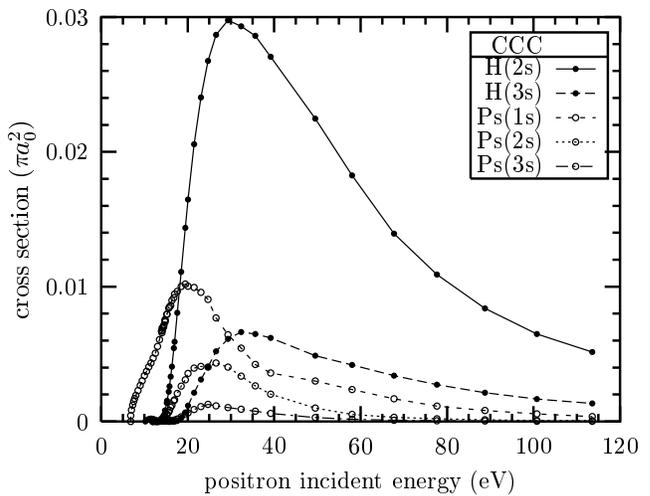}
\caption{Discrete H excitation and Ps formation cross sections for
$n\le3$ states in the S-wave
model. See text for the description of the CCC calculations.}
\label{Fig3}
\end{figure}

The excitation cross sections to $n\le3$ are given in Fig.~\ref{Fig3}. Above
the Ps(1s) threshold of 6.8~eV the Ps(1s) cross section is dominant
until around 20~eV where H(2s) becomes the largest
inelastic cross section. Most
importantly we see smooth results for all these cross sections even
though their magnitude is quite small. This gives us confidence in the 
accuracy of the pesented results generally.

In summary, the CCC method for electron-atom scattering~\cite{BS92}
has been extended to positron scattering with the inclusion of the Ps
formation channel. This channel is particularly important for the
S-wave model studied at the low and intermediate energies. Within this
model the disappearance of the Higgins--Burke type resonances is only
possible using pseudostate expansions on both the atomic and Ps
centers. Other pseudoresonances, resulting from finite pseudostate
expansions, disappear if sufficiently large expansions are taken.
Utility of using simultaneously two near-complete expansions to yield
convergent results at all energies of interest has been
demonstrated. Taking as many as 17 of both atomic- and Ps-centered states
represents some of the biggest calculations of this type ever
performed, and suggests the utility of the present numerical
implementation for full positron-atom scattering problems.  The
presented cross sections represent benchmark results that may be used
for comparison with other theories. We suspect that
CC($\overline{N},\overline{N'}$) type calculations will yield faster
convergence and greater accuracy for the full positron-atom scattering
system over the commonly-used CC($\overline{N},N'$) type calculations,
particularly for the Ps formation cross sections.

\acknowledgments
Support of the Australian Research Council and the Flinders University
of South Australia is acknowledged.  We are also indebted to the South
Australian Centre for High Performance Computing and Communications.

%\bibliographystyle{prsty}
%\bibliography{references}

\end{document}